\documentclass{Interspeech}
\usepackage{amsmath,graphicx}
\usepackage{hyperref}
\usepackage[table,xcdraw]{xcolor}
\usepackage{multirow}
\usepackage{float}                 
\usepackage{tabularx}
\usepackage{svg}
\usepackage{soul}
\usepackage{xcolor}
\usepackage{graphicx}

 \interspeechcameraready

\title{State-Space Models in Efficient Whispered and Multi-dialect Speech Recognition}

\author[affiliation={1}]{Aref}{Farhadipour}
\author[affiliation={2}]{Homayoon}{Beigi}
\author[affiliation={1}]{Volker}{Dellwo}
\author[affiliation={3}]{Hadi}{Veisi}

\affiliation{Department of Computational Linguistics}{University of Zurich}{Switzerland}
\affiliation{Recognition Technologies}{Inc. and Columbia University}{USA}
\affiliation{Faculty of New Sciences and Technologies}{University of Tehran}{Iran}
\email{aref.farhadipour@uzh.ch, beigi@recotechnologies.com, volker.dellwo@uzh.ch, h.veisi@ut.ac.ir}

\usepackage{comment}

\begin{document}

\maketitle

\begin{abstract}
    
    Whispered speech recognition presents significant challenges for conventional automatic speech recognition systems, particularly when combined with dialect variation. However, utilizing an efficient method to solve this problem using a low-range dataset and processing load is beneficial. This paper proposes a solution using a Mamba-based state-space model and four fine-tuned self-supervised models consisting of Wav2Vec2, WavLM, HuBERT, and Whisper to address the dual challenges of whispered speech and dialect diversity. Based on our knowledge, this represents the best performance reported on the wTIMIT and CHAINS datasets for whispered speech recognition. We trained the models using whispered and normal speech data across Singaporean, US, and Irish dialects. The findings demonstrated that utilizing the proposed Mamba-based model could work as a highly efficient model trained with low amounts of whispered data to simultaneously work on whispered and normal speech recognition. The code for this work is freely available \footnote{\url{https://github.com/areffarhadi/Whisper_fine_tuning_ASR}}
\footnote{\url{https://github.com/areffarhadi/Mamba-ASR}}.

\end{abstract}

\section{Introduction}
\label{sec:intro}

Unconventional or challenging speech patterns can alter the acoustic structure of speech, posing difficulties for standard automatic speech recognition systems \cite{2024gammatonegram, 2018dysarthric}. Whispered speech is a form of speech production often used voluntarily to maintain privacy or avoid disturbing others in environments such as libraries. It is sometimes produced involuntarily due to medical conditions affecting the vocal cords. This type of speech can be helpful in human-machine interaction, as individuals frequently engage in whispered speech during such interactions. However, most current speech recognition systems are designed exclusively for normal speech and thus perform poorly when recognizing whispered speech. Therefore, it is essential to adapt these systems to handle whispered speech effectively.


Just as whispering alters the acoustic structure of speech, speaking in different dialects also substantially impacts acoustic features and potentially affects the language model \cite{farhadipour2024analysis}. These variations deviate from the standardized form of a language, leading to reduced performance in speech recognition systems that are primarily trained on standard speech. As a result, systems optimized for normal speech often struggle with whispered speech or speech in unfamiliar dialects. This work aims to enhance the robustness of speech recognition models by addressing challenges related to both whispered speech and dialect variation.

Although previous studies have addressed these challenges separately, few have explored them in combination. In real-world applications, speech recognition systems must be capable of handling both whispered speech and dialect variation simultaneously. Key studies addressing whispered speech challenges include \cite{farhadipour2024leveraging} that used the WavLM model on the CHAINS dataset and achieved 9.22\% for word error rate (WER). Generative methods like recurrent neural networks and sequence-to-sequence models have been employed to handle limited data \cite{lian2019whisper}, with approaches like MelGAN and VQ-VAE used for whisper-to-speech conversion \cite{wagner2023generative}. End-to-end (E2E) systems, \cite{chang2021end}, effectively captured high-frequency structures in whispered speech, significantly reducing error rates but with a huge computational load. \cite{lin2023improving} also developed an E2E system with a WER of 36.3\%. Additionally, \cite{grozdic2017whispered} employed deep denoising autoencoders and Teager-energy-based cepstral features, leading to a 31\% increase in recognition accuracy. 

However, research on dialect variation in speech recognition is discussed in works such as \cite{li2018multi}, which explored the use of a sequence-to-sequence model, listen, attend, and spell, to train a single speech recognition system for multiple English dialects, demonstrating that incorporating dialect-specific information improves performance, outperforming individually trained models by 16.5\%. Authors in  \cite {das2021multi} investigated multi-dialect speech recognition using ensemble modeling, where dialect-specific models are combined with attention weights from an LSTM, achieving a 4.74\% WER reduction compared to a baseline, after training on 60,000 hours of speech from various English dialects.

Two strategies in designing a system for whispered speech recognition could be utilizing a self-supervised model pre-trained using a huge amount of data, and converting speech between whispered and normal styles to produce more data for training. Both of these solutions seem inefficient because they need a huge amount of data and computational resources. Whereby we proposed an efficient model for whispered speech recognition. In this paper, we used a state-space model facilitated with new mamba layers to decrease the model size and mix the small set of whispered speech in different dialects with LibriSpeech as normal speech to train the model. In this situation, we try to find an efficient way to avoid using huge self-supervised and conversion models and small datasets on the ASR designing scale. In this paper, we design several ASR systems for the English language that address the dual challenges of whispered speech and dialect variation, specifically focusing on Singaporean, US, and Irish dialects. Given the limited availability of resources for whispered speech recognition, besides the proposed Mamba-based model trained from scratch, we employed self-supervised models, consisting of Wav2Vec2 \cite{baevski2020wav2vec}, WavLM \cite{chen2022wavlm}, HuBERT \cite{hsu2021hubert}, and Whisper \cite{radford2023robust}. These state-of-the-art self-supervised models are fine-tuned on the whispered datasets to adapt to whispered speech and different dialects. However, We introduced an autoencode architecture consisting of the Mamba and ConMamba layers for the proposed state-space model. In this article, the \textit{Whisper} model developed by OpenAI is italicized to differentiate it easily from the concept of whispered speech. 

To design a multi-style and multi-dialect speech recognition model, we utilized mixed data of whisper and normal speech, using a part of the whispered portion of the wTIMIT dataset \cite{lim2010wtimit}, which features a Singaporean dialect. However, we include whispered speech samples from Irish and US dialects in the testing phase. One of the objectives of this work is to ensure the system performs well with both whispered and normal speech. Therefore, we incorporate the Irish speech dialect with the normal speaking style in the training phase.

The remainder of this paper is organized as follows: Section \ref{sec:whsp} provides an in-depth discussion of whispered speech and its characteristics. Section \ref{sec:mthd} presents the datasets and a brief overview of the proposed models. Section \ref{sec:exp} outlines the results of the proposed system, and Section \ref{sec:conc} concludes the study.

\section{whispered and Multi-dialect speech}
\label{sec:whsp}

\begin{figure}[t]
    \centering
    \includegraphics[width=0.43\textwidth, height=0.12\textheight]{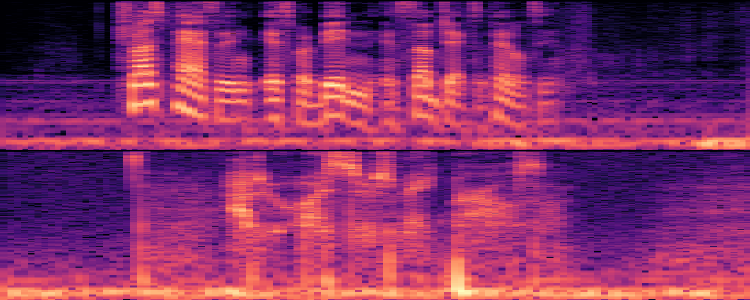}
    \caption{Differences of formats and acoustic features in normal and whispered speech from the same speaker in Spectrogram}
    \label{fig:F0}
\end{figure}

Whispered speech exhibits significant differences from normal speech in terms of its production mechanisms and acoustic features. The primary distinction lies in how they are produced. In the whispered speech, the air is exhaled through an adjusted pharynx, ensuring the vocal cords do not vibrate \cite{jovivcic2008acoustic}.

This leads to a breathier quality and a lack of the clear harmonic structures typically observed in normal speech. Consequently, formants, essential for identifying specific sounds, are less apparent in the spectrograms of whispered speech \cite{lin2023improving}. Physiological studies using magnetic resonance imaging reveal that the supraglottal structures become more constricted during a whispered speech and are positioned lower, pressing against the vocal folds to inhibit their vibration \cite{tsunoda1997laryngeal}. This production mechanism gives whispered speech its distinct hushed and breathy tone.

The elevated frequency range of whispered speech is largely due to increased airflow turbulence. This turbulence produces a noisier signal with less distinct formant structures, essential for recognizing vowels and certain consonants in regular speech. As a result, whispered speech is generally less intelligible, especially in environments with background noise. It also has a flatter spectral slope, reduced overall energy, lower amplitude, and less defined formants, particularly in lower frequencies.\ Moreover, the articulation of phonemes changes in whispered speech \cite{lin2023improving}. Studies have found that the first and second formants in whispered speech tend to shift to higher frequencies, and vowel durations are longer \cite{houle2020acoustic}.

Despite these differences, whispered speech plays a valuable role in communication, often used in settings requiring discretion, privacy, or when a quieter voice is necessary. Fig \ref{fig:F0} demonstrates the distinct differences in the spectrograms of normal and whispered speech signals produced by the same speaker. As seen in this figure, whispered speech shows lower amplitude and lacks the regular periodicity typical of normal phonation, reflecting the absence of vocal fold vibrations. The frequency content in whispered speech is more dispersed and lacks the defined harmonic patterns seen in normal speech, resulting in a more evenly distributed energy spectrum. On the other hand, normal speech exhibits higher amplitude, distinct harmonic bands, and well-defined speech formants. The spectrogram of normal speech shows brighter areas, indicating greater intensity.

However, Dialect refers to regional or social variations of a language, which, despite having many commonalities, often exhibit significant differences in various linguistic aspects. These distinctions can appear in areas such as phonology, grammar, spelling, and vocabulary \cite {das2021multi}. Consequently, automatic speech recognition (ASR) systems trained or optimized for a particular dialect tend to perform poorly when applied to a different dialect of the same language.

The combination of whispered speech and dialectal variation poses a unique challenge for ASR systems. Whispered speech alters acoustic patterns, such as the lack of vocal fold vibrations and reduced formant clarity, while dialectal differences introduce variations in phonology and vocabulary. These factors significantly reduce ASR accuracy as systems trained on standard speech or specific dialects struggle to adapt to the acoustic and linguistic changes in whispered speech across different dialects. Addressing this issue requires more advanced ASR models capable of handling the spectral differences in whispered speech and the variability of dialects.
\section{methodology}
\label{sec:mthd}
This section describes the materials and the proposed system. Specifically, we introduce the wTIMIT and CHAINS \cite{lim2010wtimit} datasets and provide a brief overview of the mamba-based model engaged with convolutions and four self-supervised models consisting of WavLM, HuBERT, Wav2Vec2, and \textit{Whisper}.

\subsection{Datasets}
\label{ssec:data}
Whispered speech datasets for the English language are limited, and the wTIMIT dataset is one of the most widely used. It contains speech data from 50 speakers, 20 with a Singaporean dialect and 30 with a US dialect. The dataset is divided into training and test sets, with both whispered and normal speech available for each dialect. In this work, Singaporean speakers were used to train the systems. So, we report the results for the Speaker-Dependent (SD) mode using Singaporean speakers and the Speaker-Independent (SI) mode using US English speakers. Consequently, in the SI condition, the dialect variation also presents an additional challenge. We separately report each dialect's performance on both whispered and normal speech.
\begin{figure}[ht]
    \centering
    \includegraphics[width=0.8\columnwidth]{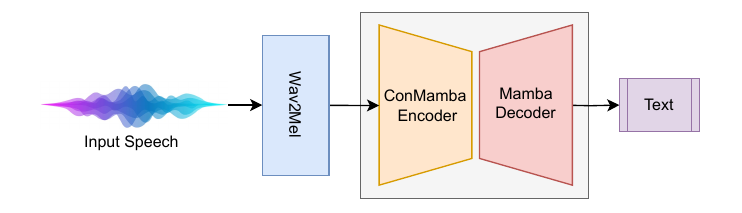}
    \caption{Proposed ConMamba-encode-Mamba-decoder ASR model}
    \label{fig:f1}
\end{figure}


The training portion of the wTIMIT dataset contains approximately 13,000 utterances and 16 hours of speech, while the test set comprises about 2,500 utterances in around 4 hours. In addition to wTIMIT, we also use the CHAINS dataset. This dataset includes solo and whispered speech from 36 speakers across two speaking styles: normal and whispered. The CHAINS dataset features 1,332 utterances per speaking style, resulting in 6 hours of speech. The speech rates in this dataset are abnormally high, which introduces an additional challenge for the models. However, We need enough data to train the proposed Mamba-based model from scratch, so we mix the 1000 hours of LibriSpeech \cite{panayotov2015librispeech} with our proposed multi-dialect multi-style training data.

\begin{table*}[t]  
\centering
\caption{Result of the baseline system as WER\% based on \textit{Whisper} model for three dialects in two speaking styles.}
\label{tab:t1}
\resizebox{0.65\textwidth}{!}{ 
    \begin{tabular}{l|ccccc}
    \rowcolor[HTML]{D0D0D0} 
    \multicolumn{1}{c|}{\cellcolor[HTML]{D0D0D0}Model} & Whisper-SG & Normal-SG & Whisper-US & Normal-US & Whisper-IRI \\ \hline
    Pre-trained Whisper & 21.67 & 7.65 & 3.4 & 5.97 & 16.58    
    \end{tabular}
}
\end{table*}

\begin{table*}[]
\centering
\caption{Result of four fine-tuned self-supervised models as WER\% based on greedy and beam searches. The greedy search wasn't used for the \textit{Whisper} model. Bolded values indicate the best result for each test set, as well as the best performance among self-supervised and Mamba models.}
\label{tab:t2}
\resizebox{0.8\textwidth}{!}{
\begin{tabular}{r|cc|cc|cc|cc|cc}
\rowcolor[HTML]{D0D0D0} 
\cellcolor[HTML]{D0D0D0}                        & \multicolumn{2}{c|}{\cellcolor[HTML]{D0D0D0}Whisper-SG} & \multicolumn{2}{c|}{\cellcolor[HTML]{D0D0D0}Normal-SG} & \multicolumn{2}{c|}{\cellcolor[HTML]{D0D0D0}Whisper-US} & \multicolumn{2}{c|}{\cellcolor[HTML]{D0D0D0}Normal-US} & \multicolumn{2}{c}{\cellcolor[HTML]{D0D0D0}Whisper-IRI} \\
\rowcolor[HTML]{D0D0D0} 
\multirow{-2}{*}{\cellcolor[HTML]{D0D0D0}Model} & Greedy                 & Beam                         & Greedy                 & Beam                         & Greedy                 & Beam                         & Greedy                 & Beam                         & Greedy                 & Beam                         \\ \hline
HuBERT                                          & 3.38                   & 2.16                         & 0.78                   & 0.70                         & 2.10                   & 1.20                         & 6.22                   & 5.79                         & 10.00                  & 9.80                         \\
Wav2Vec2                                        & 4.58                   & 1.31                         & 3.12                   & 0.92                         & 3.39                   & 2.45                         & 5.90                   & 4.28                         & 12.09                  & 11.49                        \\
WavLM                                           & 3.60                   & 0.80                         & 2.87                   & 0.63                         & 3.78                   & 2.93                         & 7.52                   & 5.99                         & 11.94                  & 12.38                        \\
Whisper                                         & -                      & \textbf{0.51}                & -                      & \textbf{0.12}                & -                      & \textbf{0.40}                & -                      & \textbf{0.92}                & -                      & \textbf{2.11}                \\
Mamba-Ver1                                           & -                      & 1.90                         & -                      & 2.19                         & -                      & 2.62                         & -                      & 1.75                         & -                      & 18.31                        \\
Mamba-Ver2                                        & -                      & \textbf{0.56}                         & -                      & \textbf{0.63}                         & -                      & \textbf{1.75}                         & -                      & \textbf{0.97}                         & -                      & \textbf{1.19}                         
\end{tabular}
}
\end{table*}


\begin{figure*}[!t]
    \centering
    \resizebox{0.7\textwidth}{!}{  
        \begin{minipage}{0.3\textwidth}
            \centering
            \includegraphics[width=\linewidth]{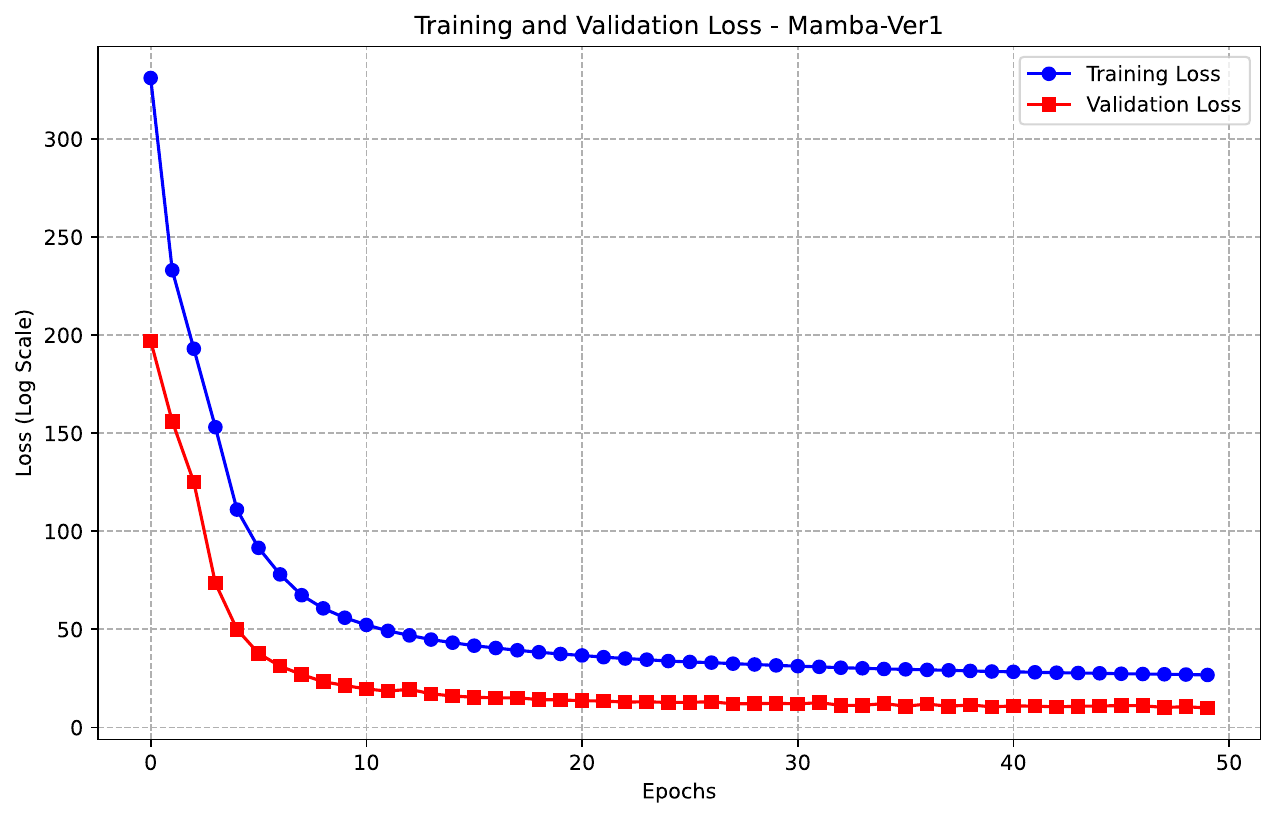}
        \end{minipage}
        \hfill
        \begin{minipage}{0.3\textwidth}
            \centering
            \includegraphics[width=\linewidth]{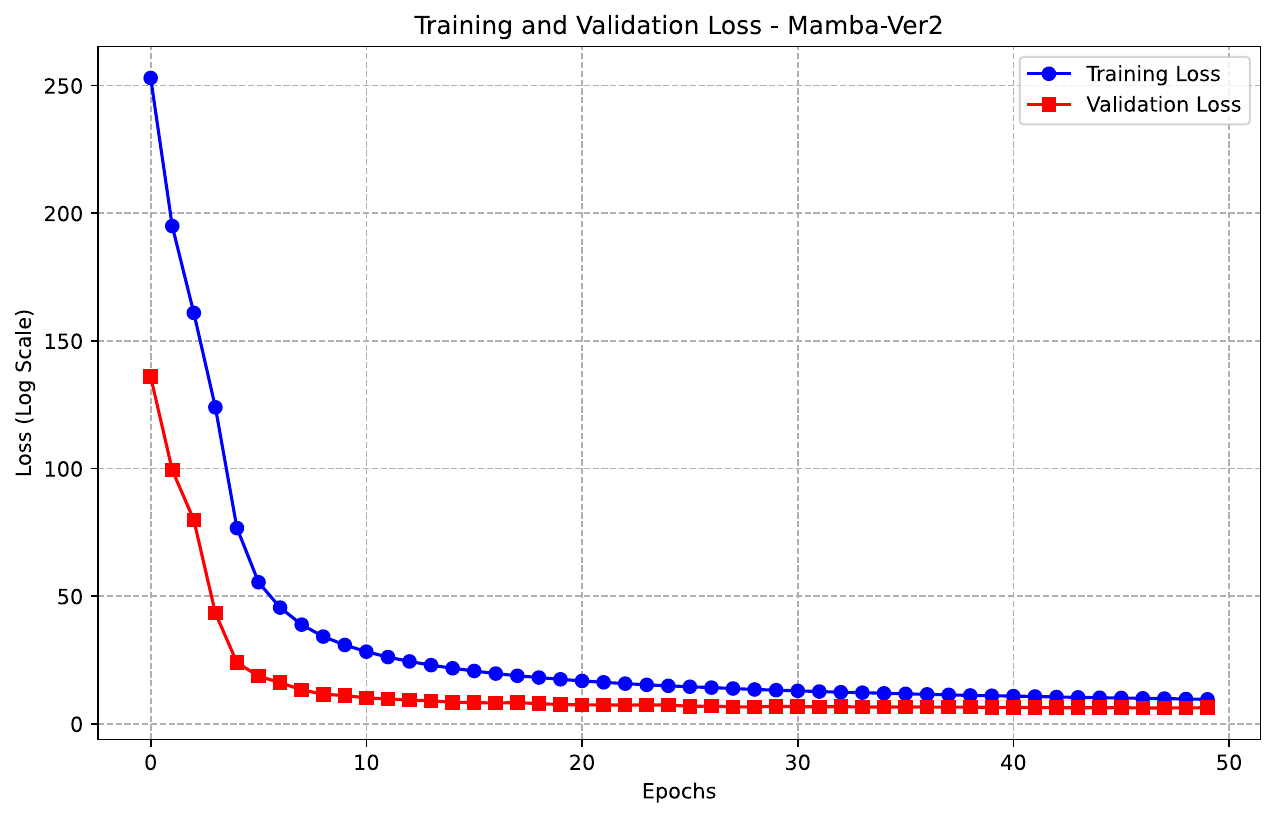}
        \end{minipage}
    }
    \caption{Training and evaluation losses for Mamba-Ver1 and Mamba-Ver2}
    \label{fig:combined}
\end{figure*}


\subsection{ConMamba Model}
\label{ssec:mamba}
 The proposed ConMamba is a speech recognition model designed as a computationally efficient alternative to transformer-based architectures, particularly Conformer. It leverages the Mamba State-Space Model (SSM) to model long-range dependencies while maintaining linear complexity in sequence length, making it more scalable than self-attention for longer speech sequences \cite{gu2023mamba}. Inspired by Conformer, ConMamba integrates depthwise-separable convolutions to enhance local feature extraction and improve phoneme boundary modeling, addressing the lack of explicit locality modeling in SSMs.

The architecture of the proposed system is illustrated in Fig \ref{fig:f1}. The ConMamba encoder features stacked bidirectional Mamba layers, each containing an SSM module that replaces self-attention with a linear selective state-space approach. Additionally, it includes a feedforward module, layer normalization, and residual connections to improve stability and enhance representation learning.

For decoding, we employ a unidirectional Mamba decoder, which processes concatenated encoder output and autoregressive token predictions using a cross-modality learning approach \cite{jiang2024speech}.
 This work uses two versions of the architecture. The Mamba-Ver1 consists of a 12-layer bidirectional ConMamba encoder with 4 heads cross attention combined with a 4-layer unidirectional Mamba decoder, operating at 25 tokens per second to ensure an optimal balance between efficiency and performance in automatic speech recognition. The Mamba-Ver2 uses the same parameters and methods in training, such as AdamW as optimizer and training for 50 epochs, but with 8 attention heads and 6 unidirectional Mamba in the decoder.

\subsection{Self-supervised Models}
\label{ssec:hub}
This work utilizes four pre-trained models widely used in various speech processing tasks. Below is a brief introduction to each model.

HuBERT is a transformer-based model that learns speech features in a self-supervised manner. During training, masked audio data is used, with k-means applied to generate discrete target labels from hidden layers. These labels serve as the predicted units for the masked sections of the input speech \cite{hsu2021hubert}.

 Wav2Vec2 has revolutionized speech processing by applying self-supervised learning directly to raw audio data. The model consists of a feature extractor built on convolutional layers and a transformer that provides a contextual understanding of the input. A masking technique is employed during training, where segments of the input speech are masked, and the network predicts the missing parts \cite{baevski2020wav2vec}.
    
WavLM is an extension of the Wav2Vec2 and HuBERT models, offering improvements in self-supervised learning. In addition to the masking technique, WavLM incorporates audio mixing, introducing a target speaker in the learning process. The model predicts masked sections while identifying speaker-specific information, making it particularly effective in tasks like speaker recognition and diarization \cite{chen2022wavlm}.

\textit{Whisper} is an encoder-decoder transformer model trained on 680,000 hours of multilingual speech. It is designed to handle various accents and background noise, making it highly versatile for speech recognition. In addition to speech recognition, \textit{Whisper} supports language identification and multilingual transcription, and it can perform tasks such as speaker recognition and voice activity detection. The decoder part uses a token set for language identification and transcription \cite{radford2023robust}.

\section{Experimental Results}
\label{sec:exp}

In this work, the proposed state-space and self-supervised models for whispered speech recognition have been trained to perform well with both whispered and normal speech. As a baseline, we evaluated the performance of the pre-trained \textit{Whisper} Large-v2 model on the test set to assess the need for a specialized system for the proposed challenges.

As previously mentioned, the \textit{Whisper} model can recognize languages and transcribe text accordingly. However, significant issues were encountered in detecting English with a Singaporean (SG) dialect despite correctly identifying the US and Irish dialects. This problem was especially pronounced in Singaporean whispered speech, where in over 90\% of cases, the language was incorrectly classified as Malay. To mitigate the impact of this language recognition error, we decoupled the multilingual components of the model by calling the tokenizer and other key modules, such as the feature extractor, separately. Table \ref{tab:t1} presents the ASR baseline results.

As shown in Table \ref{tab:t1}, while the \textit{Whisper} model achieved higher accuracy for normal speech, its performance on whispered speech was notably poorer. The WER for whispered speech was 21.67\%, 16.58\%, and 5.97\% for SG, Irish, and US dialects, respectively. Additionally, dialect-specific performance disparities were observed for normal speech, where the WER for the SG dialect was nearly twice that of the US dialect despite identical text and environmental conditions. These results emphasize the need for a system to effectively handle whispered speech and dialect variation. 
Table \ref{tab:t2} summarises the proposed systems' performance. For finetuning the self-supervised models, all parameters were kept identical across the four models for consistency in comparison. During training, we applied the SpecAugment algorithm \cite{park2019specaugment} in time-masking mode. The models were trained with a batch size of 8, using a learning rate schedule, and ran for 5 epochs to prevent overfitting. The AdamW optimizer was employed \cite{loshchilov2017decoupled}. In the decoding stage, beam search and greedy search were used for the WavLM, HuBERT, and Wav2Vec2 models, implemented using the PaddleSpeech toolbox \cite{zhang2022paddlespeech}, and results for each method are reported separately. These experiments were conducted on an Nvidia A100 80GB GPU. We used the Base+ version of WavLM, the Large-960 version of HuBERT and Wav2Vec2, and the Large-v2 version of Whisper. 

In most previous studies using the wTIMIT dataset, both dialects were included in system training. This work designed the evaluations to be more challenging and closer to real-world applications. The proposed systems were exposed to normal and whispered speech during fine-tuning for the SG dialect. In contrast, only normal speech was included for the Irish dialect, and Irish whispered speech was not encountered during training. Finally, during training, the systems were not provided with data for the US dialect, neither normal nor whispered speech.

According to the results, the proposed systems outperform the baseline \textit{Whisper} model, as seen in Table \ref{tab:t2}. The beam search algorithm consistently outperformed greedy search in most cases. Therefore, the \textit{Whisper} and Mamba models were evaluated only with beam search to reduce the computational cost. Across the five evaluation plans, the \textit{Whisper} model achieved the lowest WER. For SG normal speech, a WER of 0.51\% was obtained. For US dialect, the WER was 0.92\% for whispered speech and 0.40\% for normal speech. Finally, for whispered speech in the Irish dialect, the WER was 2.11\%. The unusually high speech rate in the CHAINS dataset for whispered speech likely contributed to the lower performance of the Irish dialect compared to other dialects.
 However, the loss value during the training process is depicted in Fig \ref{fig:combined}. The proposed Mamba-Ver2 model performed close to the \textit{Whisper} model but with a significantly smaller model size and training data. The power of this state-space model based on Mamba even could perform better than \textit{Whisper} for the whispered speech with Irish dialect with 1.19\% for WER. To design an efficient model for this challenging scenario, we can use a smaller model that we could train with tiny data compared to the \textit{Whisper} model and its massive amount of data in pre-training. 
Compared with previous works, the proposed systems with this training strategy produced near-equivalent results across various dialects and performed well on both whispered and normal speech. This demonstrates the effectiveness of the techniques employed in improving the system's robustness and ability to model different acoustical behaviors of speech signals.

\section{Conclusion}
\label{sec:conc}

Whispered speech recognition presents a significant challenge for conventional ASR systems, particularly when combined with additional factors such as limited data and dialect variation. These challenges can severely degrade the performance of traditional systems, highlighting the necessity of developing ASR models specifically tailored to handle whispered speech. However, it is very important to make an efficient system that is scalable and small enough to be usable on different sizes of computational resources. Moreover, a small set of data is needed for training and reach a reasonable result. 

This paper addressed proposed challenges with a highly efficient model based on a version of Mamba as the state-of-the-art method in modeling sequential data. On the other hand, four popular self-supervised models consisting of Wav2Vec2, WavLM, HuBERT, \textit{Whisper} were employed. We first evaluated the pre-trained Whisper model as a baseline, demonstrating that standard ASR systems struggle to perform satisfactorily on whispered speech. We fine-tuned the models to improve performance using both whispered data and diverse dialects.

The results showed that all the systems outperformed the baseline, with the \textit{Whisper} and efficient Mamba-Ver2 models delivering the best overall performance. The proposed models handled both whispered and normal speech across all three dialects (SG, US, and Irish) with high accuracy, proving their effectiveness in addressing the dual challenges of whispered speech recognition and dialect variation.

\bibliographystyle{IEEEtran}
\bibliography{template}

\end{document}